\documentstyle[aps,prl,epsf]{revtex}
\begin{document}

\newcommand{\ket}[1]{|#1\rangle}
\newcommand{\bra}[1]{\langle#1|}

\wideabs{
\title{%
	Does DLCQ S-matrix have a covariant continuum limit?
	}
\author{{\bf Masa-aki Taniguchi, Shozo Uehara, Satoshi Yamada}
	and {\bf Koichi Yamawaki}}
\address{Department of Physics, Nagoya University, Nagoya, 464-8602,
	Japan.}
\date{\today}
\maketitle
\begin{abstract}
We develop a systematic DLCQ perturbation theory and  show that DLCQ
S-matrix does not have a covariant continuum limit for processes with
$p^+=0$ exchange. This implies that the role of the zero mode is more
subtle than ever considered in DLCQ and hence must be treated with
great care also in non-perturbative approach. We also make a brief
comment on DLCQ in string theory.
\end{abstract}
\pacs{11.10.-z, 11.25.-w}
}

Discrete light-cone quantization (DLCQ) was proposed to solve the
zero-mode problem in the light-front quantization \cite{M&Y} and later
a non-perturbative calculation was carried out in field theory
\cite{P&B}. Recently the idea played an important role in
non-perturbative formulations in superstring or M-theory
\cite{BFSS,Sus} where much attention has been paid to DLCQ
\cite{foot1} in the sense of L$^3$ \cite{P&F,Sei,H&P,Bil1,U&Y}.
In contrast to string theory, local field theory in L$^3$ is
pathological, since the scattering amplitudes having zero-mode loops,
which are absent in DLCQ \cite{Yam}, diverge as $\delta(0)$ in L$^3$
\cite{H&P}.

DLCQ has a long history. In the original paper \cite{M&Y}, the
light-like coordinate $x^-$ was compactified in order to isolate the
light-cone zero modes and it was shown that the zero modes are
written in terms of the non-zero modes due to a nonlinear operator
equation (zero-mode constraint). Then the zero modes are removed from
the physical Fock space.  Because of that, DLCQ proves the popular
statement that the light-cone vacuum is trivial.

{}From the very beginning of DLCQ, however, it was known that
Poincar\'e invariance in DLCQ is not recovered as the operator algebra
in the continuum limit \cite{M&Y}, and it was further shown \cite{N&Y}
that Wightman function does not agree with the Lorentz-invariant
result in the continuum limit of DLCQ (or any other regularization
fixed to an {\it exact light-front surface}), since the dynamical
information of zero mode is lost: Namely, DLCQ has {\it no
Lorentz-invariant continuum limit} as field theory itself (operator,
Hilbert space, etc.) \cite{Yam}. The latter problem can be avoided in
the free theory if one uses {\it an explicit solution of the dynamics}
(free propagation) to regularize the theory in a way dependent on the
light-cone time.
Thus it has been expected that, if one uses explicit solution of the
dynamics, the {\it Lorentz-invariant continuum limit may be realized
on the S-matrix} which is an integrated quantity not restricted to a
fixed light-cone time \cite{Yam}.
Actually it was argued \cite{HMV} that DLCQ S-matrix has a covariant
continuum limit in the scalar field theory through a concrete
dynamics, namely the perturbation. Unfortunately, in \cite{HMV}
some diagrams having exchange of $p^+=0$ \cite{foot2} are  missing
and hence the statement is not correct.

In this paper, we develop a {\it systematic} DLCQ perturbation theory
and find that the DLCQ S-matrix disagree with a covariant one in the
continuum limit for particular processes with $p^+=0$ exchange. This
implies that zero modes in DLCQ are more subtle than ever thought even
in the perturbation. Although this discrepancy is only for the real
part of the forward scattering amplitude and hence may only occur at
zero measure part in the cross section, with no affect in total cross
section through optical theorem, it strongly suggests that similar
discrepancy may be caused by zero modes in physical quantity also in
non-perturbative calculation: The zero modes should be handled with
great care.

Let us begin with formulating the perturbative expansion (Dyson
expansion) of scalar field theory in the operator formalism.
We consider a $d$-dimensional $\lambda \phi^4$ scalar field theory
with the mass $m$ and put the system in a light-like box ($-L\leq
x^-\leq L$) with periodic boundary conditions \cite{M&Y,nota}.
We decompose the scalar field $\phi(x)$ into the zero mode
$\phi_0(x^+,x_\bot)$ $\left(\equiv (1/2L)
\int_{-L}^{L}\phi(x)dx^{-}\right)$ and the non-zero mode parts
$\varphi(x)$. Then the Hamiltonian is given by,
\begin{eqnarray}
  H&=& \int_{-L}^{L}dx^-
	\int_{-\infty}^{\infty} d^{d-2}x_{\bot}
	\left[\frac{1}{2}\varphi
	(m^2-\partial_{\bot}^2)\varphi \right.\nonumber\\
 && \left. \hspace{1cm} +  \frac{1}{2} \phi_0
	(m^2-\partial_{\bot}^2) \phi_0
       + \frac{\lambda}{4!}
	(\varphi+\phi_0)^4\right]. \label{eq:H}
\end{eqnarray}
Furthermore, from the equation of motion for $\phi_0$,
we obtain the following zero-mode constraint \cite{M&Y},
\begin{equation}
    \int^{L}_{-L}d x^- \left[(m^2- \partial_{\bot}^2)\phi_0 +
    \frac{\lambda}{3!}(\varphi+\phi_0)^3\right] = 0\,.\label{eq:ZMC}
\end{equation}
It is obvious that the zero mode is nothing but an auxiliary field
having no kinetic term.  Solving this zero-mode constraint
perturbatively (i.e., the zero mode is written in terms of the
non-zero modes perturbatively) and plugging the result into the
Hamiltonian (\ref{eq:H}), we obtain the effective interaction terms
for $\varphi$. Then using the effective Hamiltonian, we can perform
the perturbative expansion of the vacuum and scattering amplitudes
\cite{M&R,HMV,TUYY}.
In the perturbative expansion, we treat the first term in the
r.h.s. of eq.(\ref{eq:H}) as the free part of the Hamiltonian and the
rests as interactions.

\begin{figure}[htbp]
\centerline{\epsfbox{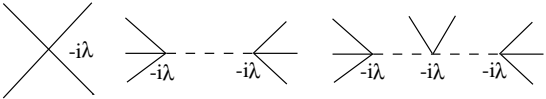}}
\caption{The effective interactions to the third order in $\lambda$.
The dotted lines represent the zero-mode propagators and the solid
lines represent the non-zero mode operators.}
\label{fig:1}
\end{figure}

Classically, or putting aside the operator ordering first, we can
straightforwardly solve the zero-mode constraint (\ref{eq:ZMC})
perturbatively and we obtain,
\begin{eqnarray}
  &&\phi_0(x^+,x_\bot)= -i\lambda \int d^dy\,
	\Delta^{xy}_{0}\,\frac{\varphi^3(y)}{3!}\nonumber \\
  &&~+(-i\lambda)^2\!\!\int\!\!d^dy\,d^dz\,\Delta^{xy}_{0}\,
    \frac{\varphi^2(y)}{2!}\,\Delta^{yz}_{0}\,\frac{\varphi^3(z)}{3!}
	+ O(\lambda^3), \label{eq:4}
\end{eqnarray}
where $\Delta^{xy}_{0}$ is the Feynman propagator for the zero mode.
\begin{eqnarray}
 \Delta^{xy}_{0} &\equiv& \frac{1}{2L} \int
	\frac{dk^- d^{d-2}k_{\bot}}{(2\pi)^{d-1}i}\,
	\frac{e^{-ik^-(x^+-y^+) +
	ik_{\bot}(x_{\bot}-y_{\bot})}}{m^2+k_{\bot}^2} \nonumber\\
 &=& \frac{1}{2iL}\,\frac{1}{m^2-\partial_{x_{\bot}}^{2}}\,
	\delta(x^+-y^+)\delta^{(d-2)}(x_{\bot}-y_{\bot}).
\end{eqnarray}
Plugging eq.(\ref{eq:4}) into the Hamiltonian (\ref{eq:H}), we obtain
the effective interaction Hamiltonian,
\begin{eqnarray}
  &&  -i\int dx^+ H_{int} = -i\lambda\int d^dx \frac{\varphi^4(x)}{4!}
	 \nonumber \\
  && \hspace{2ex}+\frac{1}{2!}(-i\lambda)^2 \int d^dx d^dy
	\left(\frac{\varphi^3(x)}{3!} \Delta^{xy}_0
	\frac{\varphi^3(y)}{3!} \right)\nonumber\\
  && \hspace{2ex}+\frac{3}{3!}(-i\lambda)^3 \int d^dx d^dy d^dz
	 \left(\frac{\varphi^3(x)}{3!}
	\Delta^{xy}_0\right. \nonumber\\
  &&	\left. \hspace{15ex}
       \times \frac{\varphi^2(y)}{2!} \Delta^{yz}_0
  \frac{\varphi^3(z)}{3!}\right) +~O(\lambda^4)\,.\label{eq:Hi}
\end{eqnarray}
The first three terms on the r.h.s. of eq.(\ref{eq:Hi})
correspond to the diagrams in Fig.\ref{fig:1}, respectively.

Now we comment on the operator ordering. In the ordinary
quantization, the equal-time commutator $[\phi(x),\phi(y)]_{x^0=y^0}$
is zero due to the micro-causality, while in the light-cone
quantization, the equal light-cone time commutator
$[\varphi(x),\varphi(y)]_{x^+=y^+}$ in general  has a non-zero value.
Thus we must fix the operator ordering in eq.(\ref{eq:Hi}).
Here we choose the Weyl ordering for the equal light-cone time
operators. Actually, it is easily checked that the Weyl ordering is
consistent with the Euler-Lagrange equation \cite{foot3}.

Having fixed the ordering of the equal light-cone time operators, we
can perform the perturbative expansion (Dyson expansion) with the
effective interaction Hamiltonian (\ref{eq:Hi}) systematically.
{}From the Schr\"odinger equation of the light-cone time evolution,
$i(\partial/\partial x^+) \ket{x^+} = H_{int}(x^+) \ket{x^+}$ in the
interaction picture, we obtain the $S$-operator,
$S = T \exp{\left[-i\int_{-\infty}^{\infty}dx^{+} H_{int}(x^{+})
\right]}$, where $T$-symbol means the light-cone time ordered product.
It should be noted that micro-causality is not necessary in order for
$T \phi(x)\phi(y) =\theta(x^+ - y^+) \phi(x) \phi(y)
+\theta (y^+ - x^+)  \phi(y) \phi(x)$ to be Lorentz-invariant,
since the sign of $x^+ - y^+$ is Lorentz-invariant wherever the two
points $x, y$ may be, in sharp contrast to the ordinary quantization
where micro-causality, $[\phi(x), \phi(y)] =0$ for $(x -y)^2 <0$, is
definitely required, since the sign of $x^0 - y^0$  is not
Lorentz-invariant in space-like region.

Since the field operators $\varphi(x)$ in the interaction picture
obey the equations of motion of the free fields, $\varphi(x)$ in
eq.(\ref{eq:Hi}) is expanded as,
\[
  \varphi(x) \!=\!\! \sum_{n >0 } \frac{1}{2L\!\cdot \!2p_n^{+}}\!\!
  \int\!\!\frac{d^{d-2}p_{\bot}}{(2\pi)^{d-2}}\!
  \left[a_n(p_{\bot})e^{-i px}\!\!+a_n^{\dagger}(p_{\bot})
	e^{i px}\right],
\]
where $p x=\{(p_{\bot}^2 +m^2)/(2 p_n^{+})\} x^+ + p_n^{+}x^-
{}-p_{\bot}x_{\bot}$, $p_n^{+}=n\pi/L$ and $\left[a_m(p_{\bot}),
a_n^{\dagger}(q_{\bot})\right]=	(2\pi)^{d-2} 2L \cdot 2p^+_n
\delta_{mn}\delta^{(d-2)}(p_{\bot}-q_{\bot})$.

Now that we have given the systematic DLCQ perturbation theory,
we can calculate the scattering amplitudes straightforwardly.
\begin{figure}[htbp]
\centerline{\epsfbox{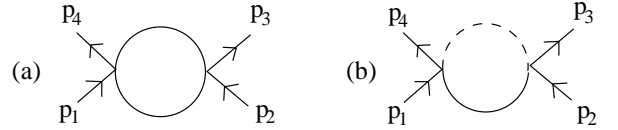}}
\caption{Examples of the scattering diagram of order $\lambda^2$.}
\label{fig:2}
\end{figure}
{}From the effective interactions in Fig.\ref{fig:1},
it is easily seen that two diagrams in Fig.\ref{fig:2} contribute
to the one-loop scattering, where external momenta are given by
$p_i=(p_{i}^-,n_i\pi/L,p_{i\bot}),\,(n_i>0,\,i=1,\cdots,4)$,
respectively. Actually we can calculate each amplitude of these
diagrams, and combine them to obtain the following scattering
amplitude $A$ \cite{foot4},
\begin{eqnarray}
  A&=& \frac{1}{2L}{\sum_{l}}' \int
	 \frac{dk^-d^{d-2}k_{\bot}}{(2\pi)^{d-1}i}
         \frac{1}{m_{\epsilon}^2-2k^-\,\frac{l \pi}{L}
	 +(k_{\bot})^2}\nonumber\\
 &&\hspace{10ex}\times\frac{1}{m_{\epsilon}^2-2\hat{k}^-\,
	 \frac{(l-n_2+n_3)\pi}{L}+(\hat{k}_{\bot})^2}\,,\label{eq:A}
\end{eqnarray}
where
\begin{equation}
	\hat{k}^-=k^--p_2^-+p_3^-,
\hspace{0.5cm}\hat{k}_{\bot}=k_{\bot}-p_{2\bot}+p_{3\bot},
\end{equation}
$m_{\epsilon}^2=m^2-i \epsilon$ and $\sum_{l}'$ stands for the
summation over $l$ without the zero-mode loop contribution
($l=0$ and $l-n_2+n_3=0$).
Notice that although the zero modes can propagate schematically as
internal lines due to the effective interactions in (\ref{eq:Hi}),
{\it we never have zero-mode loop diagrams}.

The expression of eq.(\ref{eq:A}) is not transparent when taking
the continuum limit ($L \rightarrow \infty$).
Naively, one might consider that eq.(\ref{eq:A}) would agree with the
covariant answer in the continuum limit \cite{HMV}. However, in DLCQ,
{\it the continuum limit should be taken after the whole calculations
are done with finite $L$. Hence we first perform the momentum
integrals in (\ref{eq:A}).}  Using the usual parameter integral
formula, $[X-i\epsilon]^{-1} = i \int_0^{\infty}d\alpha\,
e^{-i\alpha(X-i\epsilon)}$, we obtain,
\begin{eqnarray*}
  && A =\frac{i}{2L} {\sum_{l}}'
	\int \frac{ dk^- d^{d-2}k_{\bot}}{(2\pi)^{d-1}}\,
	\int_0^{\infty}d\alpha_1 d\alpha_2\,\\
 && \times \exp{\left[-i\alpha_1 \left\{ m_{\epsilon}^2-2k^
	-\frac{l\pi}{L}	+(k_{\bot})^2 \right\} \right]}\\
 && \times \exp{\left[-i\alpha_2 \left\{ m_{\epsilon}^2 -2\hat{k}^-
	\frac{(l-n_2+n_3)\pi}{L} +(\hat{k}_{\bot})^2
	\right\}\right]}.
\end{eqnarray*}
Then, the integral over $k^-$ gives a $\delta$-function,
\begin{eqnarray*}
  A&=&\frac{i}{4\pi}{\sum_{l}}'\int
	\frac{d^{d-2}\tilde{k}_{\bot}}{(2\pi)^{d-2}}
	\int_0^{\infty}d\alpha \int_0^{1}d\gamma\,
	\delta\bbox(l-\gamma(n_2-n_3)\bbox)\nonumber\\
  &\times&\exp\left[-i\alpha\{m_{\epsilon}^2
	+(\tilde{k}_{\bot})^2 -
	\gamma(1-\gamma)(p_2-p_3)^2 \}\right],
\end{eqnarray*}
where we have changed the variables
$(k_\bot,\alpha_1,\alpha_2)\rightarrow
(\tilde{k}_{\bot},\alpha,\gamma)\equiv
(k_{\bot}+\gamma(p_{3\bot}-p_{2\bot}),\alpha_1+\alpha_2,
\alpha_2/(\alpha_1+\alpha_2))$. Wick-rotating the variable $\alpha
\rightarrow -i \alpha$, we finally get,
\begin{eqnarray}
  A&=& \frac{1}{2^d\pi^{\frac{d}{2}}}{\sum_{l}}'
    \int_0^{\infty}d\alpha\,\alpha^{\frac{4-d}{2}}
    \int_0^{1}d\gamma\,\delta\bbox(\alpha(l-\gamma(n_2-n_3))\bbox)
    \nonumber\\
  &&\times\exp\left[-\alpha\{m^2
	-\gamma(1-\gamma)(p_2-p_3)^2\}\right]\label{eq:9}\,.
\end{eqnarray}
One comment is in order: If eq.(\ref{eq:9}) had included a zero-mode
loop contribution ($l=0,n_2=n_3$), which is actually absent in
${\sum_{l}}'$, it would suffer from a pathological divergence of
$\delta(0)$. This is, in fact, the divergence in ref.\cite{H&P} in
L$^3$. In DLCQ, however, this problem is absent from the outset
\cite{Yam}.

Now we are ready to examine the continuum limit
($L\rightarrow\infty$). First, we consider the $n_2 \neq n_3$ case in
eq.(\ref{eq:9}). Performing the $\gamma$-integration in
eq.(\ref{eq:9}), we have \cite{Bil3},
\begin{eqnarray}
  A&=& \frac{1}{2^d\pi^{\frac{d}{2}}}
	\int_0^{\infty}d\alpha\,\alpha^{\frac{2-d}{2}}
        e^{-\alpha m^2} \frac{1}{|n_2-n_3|} \label{44'}\\ \nonumber
       &&  \times \left\{1 + \sum_{l=1}^{|n_2-n_3|-1}
          \exp\left[\alpha
	\gamma_l\left(1-\gamma_l\right)
         (p_2-p_3)^2\right] \right\}\,,
\end{eqnarray}
where $\gamma_l=l/|n_2-n_3|$ and we have used the prescription
$\int_0^a dx\delta(x) f(x)=(1/2)f(0)$.
The first term ``1'' in the brace is the sum of $l=0$ and $l=|n_2-n_3|$
contributions which correspond to the diagram having a zero-mode
propagator (Fig.\ref{fig:2}(b)).
Now it is easy to take the continuum limit, $|n_2-n_3| \rightarrow
\infty$, of eq.(\ref{44'}). In fact, we obtain,
\begin{eqnarray}
 A=\frac{1}{2^d\pi^{\frac{d}{2}}}
    \int_0^{\infty}d\alpha\,\alpha^{\frac{2-d}{2}}\int_0^{1}d\gamma~
    e^{-\alpha\{m^2 -\gamma(1-\gamma)(p_2-p_3)^2\}}~,\nonumber
\end{eqnarray}
which coincides with the covariant result.
Note that in the continuum limit, the contributions of the diagram in
Fig.\ref{fig:2}(b) vanish ($1/|n_2-n_3|\rightarrow0$) \cite{HMV}.

Next we consider the $n_2=n_3$ case in eq.(\ref{eq:9}) \cite{foot2}.
Such a case is precisely what was missing in \cite{HMV} and
actually yields non-covariant continuum limit.
For clarity, we consider a diagram in two dimensions ($d=2$)
\cite{foot5} where the external momenta $p_i\,\,(i=1, \cdots,4)$
satisfy $p_1=p_4$ and $p_2=p_3$ in Fig.\ref{fig:2}(a), i.e., the
forward scattering.
Note that there is no $p^+$-momentum exchange in this process and
hence the diagram of Fig.\ref{fig:2}(b) does not exist.
{}From eq.(\ref{eq:9}), the amplitude $A_F$ is given by,
\begin{equation}
  A_F= \frac{2}{2^2\pi}\sum_{l>0}\frac{1}{l}
	\int_0^{\infty}d\alpha\,\alpha\delta(\alpha)\,
	\exp(-\alpha m^2)\,.
\end{equation}
Furthermore, changing the variable $\alpha\rightarrow\beta=2
(l\pi/L)\alpha$ for later convenience, we obtain,
\begin{equation}
  A_F = \frac{1}{2L}\sum_{l>0}\frac{1}{2(\frac{l\pi}{L})^2}
	\int_0^{\infty}d\beta\,\beta\,\delta(\beta)\,
	\exp\left(-\frac{\beta m^2}{2\frac{l\pi}{L}}\right)
	\label{eq:11}.
\end{equation}
This becomes zero due to
$\int_0^{\infty}d\beta\beta\delta(\beta)f(\beta)=0~$.
Thus the scattering amplitude of this process is zero in the
continuum limit ($L\rightarrow\infty$).

On the contrary, the covariant amplitude of the same process
$A_F^{cov}$ is given by,
\begin{eqnarray}
  A_F^{cov}&=&\int \frac{dk^0dk^1}{(2\pi)^2i}
   \left(\frac{1}{m_{\epsilon}^2-(k^0)^2+(k^1)^2}\right)^2 \nonumber\\
  &=& \frac{1}{2^2\pi}\int_0^{\infty} d\alpha \int_0^1 d\gamma\,
    e^{-m^2\alpha }=\frac{1}{4\pi m^2}\label{eq:Ac}\,.
\end{eqnarray}
This is non-zero and finite! The continuum limit of $A_F\,(=0)$ does
not coincide with $A_F^{cov}$.

To see why this discrepancy has occurred, we consider the continuum
light-front amplitude of the same process. The amplitude $A_F^{LC}$ is
given by,
\begin{equation}
  A_F^{LC}= \int \frac{dk^+dk^-}{(2\pi)^2i}\left(
	\frac{1}{m_{\epsilon}^2 - 2k^+k^-}\right)^2\,.
\end{equation}
Similarly to the above, we obtain,
\begin{equation}
  A_F^{LC}   =\int_{0}^{\infty}\frac{dk^+}{4\pi(k^+)^2}
	\int_0^{\infty}d\beta\,\beta\,\delta(\beta)\,
	\exp\left(-\frac{\beta m^2}{2k^+}\right)\label{eq:Al}.
\end{equation}
Note that if we discretize the light-cone coordinate,
$k^+\rightarrow (l\pi/L)$ and
$(1/2\pi)\int_0^{\infty}dk^+\rightarrow (1/2L)\sum_{l>0}$,
then (\ref{eq:Al}) becomes (\ref{eq:11}).
At first sight, (\ref{eq:Al}) seems to be zero similarly to
(\ref{eq:11}) due to
$\int_0^{\infty}d\beta\beta\delta(\beta)f(\beta)=0$.
However, this is not the case. The reason is that once we put
$\int_0^{\infty}d\beta\beta\delta(\beta)\exp\left(-\beta
m^2/2k^+\right)$ in eq.(\ref{eq:Al}) zero, the remaining integral
$\int_{0}^{\infty}dk^+[4\pi(k^+)^2]^{-1}$ diverges, i.e.,
we shall suffer from $\infty\times0$ in such a calculation of
(\ref{eq:Al}). The proper procedure is, in fact, to perform
$k^+$-integral first and then $\beta$-integral afterward.
Changing the variable $k^+\rightarrow k=1/k^+$, we can easily carry
out the procedure,
\begin{eqnarray}
  A_F^{LC}&=& \int_0^{\infty} d\beta\,\beta\,\delta(\beta)
	\int_{0}^{\infty}\frac{dk}{4\pi}\,
	\exp\left(-\frac{\beta m^2 k}{2}\right)\nonumber\\
  &=&\frac{1}{2\pi m^2}\int_0^{\infty}
	d\beta\,\delta(\beta) =\frac{1}{4\pi m^2}\,,
\end{eqnarray}
where we have used the prescription
$\int_0^{\infty}d\beta\delta(\beta)f(\beta) =(1/2) f(0)$ as before.
This exactly coincides with the covariant result (\ref{eq:Ac}).

It is now clear why the discrepancy between $A_F$ and $A_F^{cov}$ has
occurred even in the continuum limit. In contrast to eq.(\ref{eq:Al}),
since $(1/2L)\sum_{l>0}[2(l\pi/L)^2]^{-1} = L/24$ is finite, we can
put $\int_0^{\infty}d\beta\beta\delta(\beta) \exp\left(-\beta
m^2/(2l\pi/L)\right )=0$ in (\ref{eq:11}), and hence we have $A_F=0$.
In DLCQ, since the continuum limit should be taken after the whole
calculations are done with finite $L$, we cannot obtain any non-zero
result in eq.(\ref{eq:11}) even in the continuum limit and the
covariant result (\ref{eq:Ac}) can never be reproduced in DLCQ.
Thus the continuum limit of the DLCQ S-matrix does not coincide with
the covariant one in the specific kinematics where there's no $p^+$
exchange. This means that even differential cross sections in DLCQ
differ from the covariant ones in the continuum limit.

However, since in the integral over the scattering angle, the region
which contributes to the discrepancy has measure zero, the discrepancy
does not affect total cross sections and the experimentally measurable
values of differential cross sections, which are actually integrated
in some finite region over the scattering angle. Besides, the forward
scattering amplitude is related to the total cross section due to the
optical theorem, however, the conflict part does not contribute to the
imaginary part of the forward scattering amplitude and hence the total
cross section remains the same.

Though the region of breaking Lorentz invariance of S-matrix has
measure zero in the configuration space of external momenta, this
explicit result nevertheless implies that {\it the role of the zero
modes is more subtle than ever thought in DLCQ.} Thus various DLCQ
calculations, perturbative or non-perturbative, should be done with
great care on the zero modes.

Finally, we make a brief comment on DLCQ in string theory.
Due to the s-t channel duality, a string scattering diagram of a
certain order in string perturbation contains all the possible field
theoretical Feynman diagrams of the same order \cite{U&Y}, which wipes
off such a discrepancy appeared in field theory.


We would like to thank T.~Kugo for useful discussions. KY is supported
by the Grant-in-Aid for Scientific Research (B) \#11695030 and
(A) \#12014206).


\begin{thebibliography}{99}
\bibitem{foot1} The meaning of ``DLCQ'' is sometimes confusing. In
	this paper we use ``DLCQ'' as a quantization on the exact
	light-front surface (i.e., conventional DLCQ \cite{M&Y,P&B})
	and ``Light-Like Limit
	(L$^3$)'' as a procedure to quantize on an almost light-front
	surface and then to take a limit by which the surface becomes
	the exact light-front one.
\bibitem{foot2} The kinematics, where the light-cone momentum $p^+$
	is not exchanged, is very important in the context of Matrix
	theory. In fact, as a first step, the tree-level processes in
	this kinematics have been considered intensively,
	and the remarkable agreements
	between	the potentials in finite $N$ Matrix theory and those
	in DLCQ supergravity have been reported \cite{Tay}.
\bibitem{nota} In this paper, we use a convention of the
	light-cone coordinates: $x^{\mu} =(x^+,x^-, x_{\bot})$, where
	$x^{\pm}=(1/\sqrt2) (x^{0}\pm x^1)$, $x_{\bot} =(x^2, \cdots,
	x^{d-1})$.  We treat $x^+$ as a light-cone time.
\bibitem{foot3} In principle, we can choose
	other orderings, e.g., normal ordering. If we had chosen the
	normal ordering in eq.(\ref{eq:Hi}), the diagram (b) in
	Fig.\ref{fig:2} would be absent. However this does not
	alter our conclusion \cite{TUYY}.
\bibitem{foot4} In eq.(\ref{eq:A}), though we omit the symmetry factor
	and the coupling $\lambda$, they are the same as the covariant
	ones.
\bibitem{foot5} In this paper, we consider only the process in two
	dimensions. However, similarly to the $d=2$ case, we can show
	easily $A_F=0$ in $d<4$. Furthermore, though the amplitude
	$A_F$ diverges in $d=4$, we can show that the regularized
	amplitude becomes zero in the well-known regularization methods
	\cite{TUYY}.
\end{thebibliography}
\end{document}